\documentclass[aps,prx,floatfix,twocolumn,longbibliography,nofootinbib]{revtex4-2}

\usepackage[utf8]{inputenc}
\usepackage[T1]{fontenc}
\usepackage{physics}
\usepackage{graphicx}
\usepackage{amsmath, amsfonts, amssymb}
\usepackage{bbm, bm}
\usepackage{comment}

\usepackage[breaklinks=true, colorlinks]{hyperref}
\hypersetup{linkcolor=red, citecolor=blue}

\usepackage[dvipsnames]{xcolor}
\usepackage{makecell}
\usepackage{upgreek}
\usepackage{nicefrac}

\usepackage{tikz}
\usepackage{qcircuit}
\usetikzlibrary{calc}

\newcommand{\spinchain}[3]
{%
  \begin{scope}[shift={#1},scale=0.25]
    \foreach[count=\j]\i in {#3}
    {%
      \coordinate (#2\j) at (\j,-1);
      \ifnum\i = 1
        \draw[-stealth,red,line width= 0.4mm] (\j,-1) --++ (0,2);
      \else\ifnum\i = -1
        \draw[-stealth,blue,line width= 0.4mm] (\j,1) --++ (0,-2);
      \else
        \draw[thick,green,fill=white] (\j,0) circle (0.16);
     \fi\fi
    }
  \end{scope}
}

\begin{document}

\title{Quantum enhanced Markov chains require fine-tuned quenches}

\author{Alev Orfi}
\affiliation{Center for Computational Quantum Physics, Flatiron Institute, 162 5th Avenue, New York, NY 10010, USA}
\affiliation{Center for Quantum Phenomena, Department of Physics, New York University, 726 Broadway, New York, New York 10003, USA}

\author{Dries Sels}
\affiliation{Center for Computational Quantum Physics, Flatiron Institute, 162 5th Avenue, New York, NY 10010, USA}
\affiliation{Center for Quantum Phenomena, Department of Physics, New York University, 726 Broadway, New York, New York 10003, USA}

\date{\today}

\begin{abstract}
Quantum-enhanced Markov chain Monte Carlo, an algorithm in which configurations are proposed through a measured quantum quench and accepted or rejected by a classical algorithm, has been proposed as a possible method for robust quantum speedup on imperfect quantum devices. While this procedure is resilient to noise and control imperfections, the potential for quantum advantage is unclear. By upper-bounding the gap of the Markov chain, we identify competing factors that limit the algorithm's performance. One needs the quantum dynamics to efficiently delocalize the system over a range of classical states, however, it is also detrimental to introduce too much entropy through the quench. Specifically, we show that in the long-time limit, the gap of the Markov chain is bounded by the inverse participation ratio of the classical states in the eigenstate basis, showing there is no advantage when quenching to an ergodic system. For the paradigmatic Sherrington-Kirkpatrick and 3-spin model, we identify the regime of optimal spectral gap scaling and link it to the system's eigenstate properties.   
\end{abstract}

\maketitle

\section{Introduction}
\label{sec:introduction}

Sampling from a Boltzmann distribution of a classical system is a critical task across multiple fields. Consider a classical Hamiltonian, $H_c$, which characterizes a system of $N$ binary random variables $x_i = \pm 1$. The Boltzmann distribution, 
\begin{equation}\label{eq:boltzmann}
\pi(x) = \frac{1}{\mathcal{Z}}e^{-\beta H_c(x)},
\end{equation}
describes the probability of the system occupying the configuration $x=\{x_i\}$ at a specified temperature $T=1/\beta$. As the partition function, $\mathcal{Z}$, is typically intractable, sampling from the Boltzmann distribution is often the only approach for estimating the system's thermodynamic properties. This class of sampling problem is additionally a common subroutine in machine learning and combinatorial optimization, often emerging as a computational bottleneck in these methods.

Several quantum algorithms for this task, such as quantum walks and quantum simulated annealing, are known to give algorithmic speedups over classical methods for certain problems~\cite{Szegedy,somma2008quantum,wocjan2008speedup,poulin2009sampling,bilgin2010preparing, temme2011quantum,yung2012quantum,montanaro2015quantum,chowdhury2016quantum,harrow2020adaptive, lemieux2020efficient,arunachalam2022simpler,rall2023thermal,chen2023efficient}. However, these algorithms rely on fault-tolerant quantum computers with many qubits, making them infeasible on today's devices. Current quantum devices are limited in size and performance, and few algorithms for sampling from a Boltzmann distribution are available within the constraints of current hardware~\cite{wild2021quantum,wild2021long,layden2023quantum,zhang2023dissipative,ding2308single}. 

One such algorithm, introduced in Ref.~\cite{layden2023quantum}, the quantum-enhanced Markov chain Monte Carlo, demonstrates promising error resilience. This hybrid approach constructs a Markov chain over classical configurations, proposing new configurations based on measured quantum evolution. Next, a classical step decides whether to accept or reject the proposed configuration as the new state of the chain. Empirical tests reported in Ref.~\cite{layden2023quantum} indicated a polynomial speedup in sampling low-temperature spin glass problems compared to uniform and local classical MCMC methods. Additionally, this algorithm was as a subroutine in an adversarial quantum auto-encoder, showing a large temperature region with a larger spectral gap compared to its classical counterparts \cite{AQAM}. In Ref.~\cite{orfi2024bounding}, it was shown that the speedup is not generic across models as it is provably absent for some worst-case problems. In this work, we extend the analysis to a more practically relevant class of Hamiltonians and show generic performance behaviour related to the ergodicity of the quantum quench, allowing for the identification of favourable performance regimes.

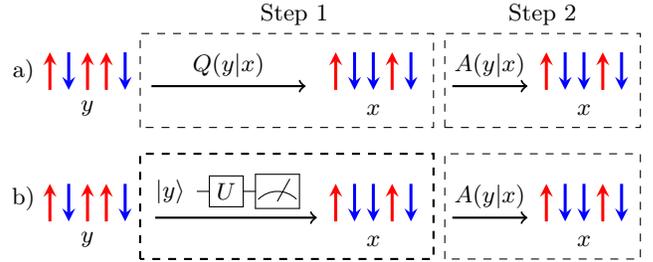
\begin{figure}
    \centering
    \begin{tikzpicture}
        \node at (0,0) {a)};
        \node[above] at (0.85,-0.7) {$y$};
        \spinchain{(0.1,0)}{chain1}{1,-1,1,1,-1}; 
        \spinchain{(3.9,0)}{chain2}{1,-1,-1,1,-1}
        \spinchain{(6.7,0)}{chain3}{1,-1,-1,1,-1}
        \draw[->, thick] (1.7,-0.2) -- (3.75,-0.2)node[midway, above] {$Q(y|x)$ };
        \draw[->, thick] (5.7,-0.2) -- (6.7,-0.2)node[midway, above] {$A(y|x)$};
        \draw[dashed] (1.55,-0.75) rectangle (5.45,0.5);
        \node[above] at (4.65,-0.7) {$x$};
        \draw[dashed] (5.6,-0.75) rectangle (8.2,0.5);
        \node[above] at (7.45,-0.7) {$x$};
        \node[above] at (3.6,0.5) {Step 1};
        \node[above] at (6.9,0.5) {Step 2};

        \node at (0,-1.7) {b)};
        \node[above] at (0.85,-2.45) {$y$};
        \spinchain{(0.1,-1.75)}{chain1}{1,-1,1,1,-1}; 
        \spinchain{(3.9,-1.75)}{chain2}{1,-1,-1,1,-1}
        \spinchain{(6.7,-1.75)}{chain3}{1,-1,-1,1,-1}
        \draw[->, thick] (1.75,-1.95) -- (3.9,-1.95)node[midway, above,, xshift=0.5em] {\Qcircuit @C=0.5em @R=.5em {\lstick{\ket{y}} &\gate{U} &\meter }} ;
        \draw[->, thick] (5.7,-1.95) -- (6.7,-1.95)node[midway, above] {$A(y|x)$};
        \draw[dashed, line width=0.8pt] (1.55,-2.5) rectangle (5.45,-1.1);
        \node[above] at (4.65,-2.45) {$x$};
        \draw[dashed] (5.6,-2.5) rectangle (8.2,-1.1);
        \node[above] at (7.45,-2.45) {$x$};
    \end{tikzpicture}
    \caption{Illustrated in (a) is the conventional two-step Markov chain update strategy. In contrast, the quantum-enhanced Markov chain Monte Carlo algorithm alters only the initial step of this process, as depicted in (b).}
    \label{fig:stepDiagram}
\end{figure}

\section{Algorithm Overview}
Markov chain Monte Carlo (MCMC) is an extremely successful and widely used class of sampling algorithms. These methods rely on Markov chains designed to equilibrate to the desired probability distribution. Often, these chains are constructed with a proposal step followed by an acceptance step, facilitating easier engineering to ensure the conditions for convergence are met. Specifically, if the chain is in the state $y$, a new state $x$ is proposed with probability $Q(x|y)$. In the acceptance step, this new state is adopted as the new state of the chain, with probability $A(x|y)$, known as the acceptance probability. This process is outlined in Fig.~\ref{fig:stepDiagram}(a).

The quantum-enhanced Markov chain Monte Carlo algorithm follows this two-step procedure, where only the proposal step requires quantum computation, as seen in Fig.~\ref{fig:stepDiagram}(b). The current state of the Markov chain, a classical configuration, is encoded as a computational basis state $\ket{x}$. A unitary $U$ is applied to the state and then the system is measured in the computational basis, resulting in a new classical configuration. This method results in the following proposal probability, 
\begin{equation}
    Q(x|y) = |\langle x|U|y\rangle|^2.
\end{equation}
A common acceptance rule, the Metropolis-Hasting probability, 
\begin{equation}\label{eq:HMacceptance}
    A(x|y)  = \min\left(1,e^{-\beta(H_c(x)-H_c(y))}\frac{Q(y|x)}{Q(x|y)}\right),
\end{equation}
depends only on the ratio of proposal probabilities. If $Q(y|x)=Q(x|y)$, the acceptance probability can be computed efficiently even if the proposal probability is intractable. With this construction, the chain satisfies the detailed balance condition. If $Q(x|y)>0$ $ \forall x,y$, the chain is aperiodic and irreducible; ensuring convergence to the desired distribution \cite{layden2023quantum}. 

Due to the classical acceptance step, this algorithm maintains convergence even with errors in the proposal step, making it well-suited for current noisy quantum devices. The only problematic errors are those that remove the needed proposal probability symmetry. However, mitigation methods such as state preparation and measurement twirling can be employed \cite{layden2023quantum}. Although many choices of $U$ meet the symmetry requirement, they do not all lead to an algorithmic improvement over classical methods. In this work the focus is on evolution of the form $U=e^{-iHt}$ with the following time-independent Hamiltonian, 
\begin{equation}\label{eq:ham}
    H = H_c + h\sum_{i=1}^N \sigma^x_i.
\end{equation}

After initialization, Markov chains require a warm-in period, known as the mixing time, to equilibrate to its stationary distribution. Specifically, the mixing time is the number of steps required for the total variation distance to $\pi$ to be $\epsilon$-small. On difficult problems, chains suffer from slow convergence, making the mixing time the important metric in MCMC performance. 

A Markov chain can be specified with a stochastic transition matrix $P$ whose elements $P(y,x)$ describe the probability of moving from the state $y$ to $x$. For the two-step construction, 
\begin{equation}\label{eq:Pconstruct}
    P(y,x) = \begin{cases}
     Q(x|y)A(x|y), & \text{if}\ x\neq y\\
     1-\sum_{z\neq x} Q(x|z)A(x|z), & \text{if}\ x=y
    \end{cases}.
\end{equation}
As $P$ is stochastic, its eigenvalues have a magnitude at most 1, and the unique stationary distribution is associated with the eigenvalue 1. The repeated application of $P$ determines the distribution of the chain, therefore, the mixing time can be connected to the spectral gap $\delta = 1 -|\lambda_2|$ of $P$, formally: 
\begin{equation}\label{eq:mixingTimeBound}
   (\delta^{-1}-1)\ln\left(\frac{1}{2\epsilon}\right)\leq t_{\text{mix}} \leq \delta^{-1}\ln\left(\frac{1}{\epsilon \pi_{\min}}\right)
\end{equation}
where $\pi_{\min} = \min_{x\in S}\pi(x)$ \cite{levin_MarkovChainsMixingTime}. The spectral gap is commonly used in performance analyses of MCMC algorithms. Exact spectral gap calculation is, however, limited to small systems as the transition matrix grows with the dimension of the state space and quickly becomes intractable to diagonalize. 

The performance of the quantum-enhanced Markov chain Monte Carlo algorithm was probed through spectral gap calculations on the Sherrington-Kirkpatrick (SK) model in Ref.~\cite{layden2023quantum}. The two free parameters of the proposal strategy, $t$ and $h$, were randomly drawn at each MCMC step. This strategy was compared to local and uniform classical proposal strategies. Even without optimizing the free parameters, the algorithm exhibited favourable mixing times in the low-temperature regime over the classical counterparts. Although these numerical studies indicated a polynomial speedup, they are limited to small systems. It is unclear whether this advantage persists for larger systems, given the significant finite-sized effects known in the SK model.

The improvement in the mixing time is suggested to be due to the proposal strategy favouring states close in energy, as the quantum evolution depends on the classical Hamiltonian. These states can be difficult to propose classically as they can be decorrelated from the initial state due to the thermalization of the system after the quantum quench. Here, we focus on how the mixing time depends on the free parameters $t$ and $h$ as a strategy to probe the underlying mechanism responsible for the found advantage. Understanding this dependence is essential in knowing if a similar improvement can be obtained where this quantum evolution is classically simulatable. Through analysis of the bottlenecks of the quantum-enhanced method, we show that the speedup observed numerically can only be achieved when quenching to a non-ergodic quantum system, highlighting the difficulty in achieving a generic speedup.

\section{Bottleneck Analysis}
The algorithmic performance of MCMC methods is often not formally derived but determined empirically. This poses a difficulty for the quantum-enhanced method as very limited system sizes are accessible. However, rigorous upper bounds for the mixing time of classical MCMC methods have been found in certain cases using analytical techniques \cite{levin_MarkovChainsMixingTime}. One such technique is to analyze the bottlenecks of the chain, which restrict the chain's movement from one section of states to another. 

\begin{figure}
    \centering
    \begin{tikzpicture}[scale=1.1] 
        \tikzset{mynode/.style={circle, draw, minimum size=5mm, inner sep=0pt, text centered}}
        \node[mynode] (x) at (0,0) {$x$};
        \node[mynode] (y) at (2,1) {$y$};
        \node[mynode] (a) at (3,1.2) {};
        \node[mynode] (b) at (1.6,-0.1) {};
        \node[mynode] (c) at (1.4,-1) {};
        \node[mynode] (d) at (2.8,-0.6) {};
        \node[mynode] (e) at (3.6,0.3) {};
        \node[mynode] (f) at (3.7,-1.5) {};
        \node[mynode] (g) at (4.7,-0.9) {};
        \node[mynode] (h) at (5,-2) {};
        \node[mynode] (i) at (6,-0.5) {};
        \node[mynode] (j) at (4.3,-2.5) {};
        \node[mynode] (k) at (6.2,-2) {};
        \node[mynode] (l) at (5.2,-3) {};
        \foreach \source/\dest in {x/y,y/a,y/b,x/b,b/c,b/a,y/d,b/d,x/c,a/e,d/e,e/b,e/f,c/f,f/g,f/h,g/i,i/h,h/g,g/j,j/h,h/k,l/j,k/g}
            \draw (\source) -- (\dest);
    
        \draw[dashed] (1,-2.5) -- (6,1);
        \node[above, rotate=27] at (1, 0.6) {$\pi(x)P(x,y)$};
    \end{tikzpicture}
    \caption{Equilibrium flow graph of an MCMC method. The nodes represent different spin configurations, and the edge weights denote the equilibrium flow between these configurations. The normalized flow through the most restricted graph cut, shown with the dashed line, provides an upper bound for the spectral gap, as expressed in Eq.~\eqref{eq:upperbound}.}
    \label{fig:bottleneck_graph}
\end{figure}
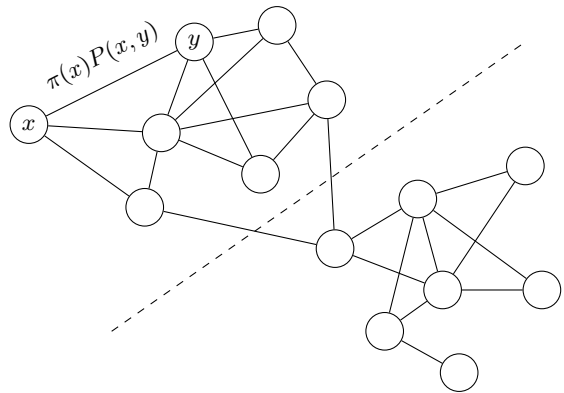

The equilibrium flow between two configurations $x$ and $y$ is defined as $E(x,y)=\pi(x)P(x,y)$. There is a natural graph corresponding to any MCMC method, depicted in Fig.~\ref{fig:bottleneck_graph}, where the nodes represent the possible configurations and the edge weights indicate the equilibrium flow between these configurations. The mixing time of the chain is related to the equilibrium flow between a subset of configurations $S$, to its complement,
\begin{equation}
    E(S,S^c) = \sum_{x\in S, y\in S^c}\pi(x)P(x,y).
\end{equation}
An upper bound can be placed on the spectral gap through minimization over possible subsets \cite{levin_MarkovChainsMixingTime}, 
\begin{equation}\label{eq:upperbound}
    \delta \leq \min_{S: \pi(S)\leq 1/2} \frac{E(S,S^c)}{\pi(S)\pi(S^c)} := \Lambda_{\text{min}},
\end{equation}
where $\pi(S)=\sum_{x\in S} \pi(x)$ is the Boltzmann weight of the subset $S$. Here, the minimization can be interpreted as finding the cut with the most restricted equilibrium flow, thus determining the chain's mixing time. 

At high temperatures, the Boltzmann distribution is close to uniform and, therefore, easy to sample. The difficulties arise at lower temperatures, where the distribution favours low-energy configurations. We will focus on this low-temperature regime, where the quantum-enhanced method numerically showed favourable mixing time. The equilibrium flow of any set $B$ to its complement provides a further upper bound on the spectral gap of this chain,
\begin{equation}\label{eq:lambdaB}
    \delta \leq  \Lambda_{\text{min}} \leq \Lambda(B)  =  \frac{E(B,B^c)}{\pi(B)\pi(B^c)}.
\end{equation}
In the low-temperature regime, for some models, one can devise a reasonable choice of $B$ resulting in a tight bound on the spectral gap. This method allows for the behaviour of the MCMC to be probed for system sizes where exact spectral gap calculation is not accessible numerically \cite{orfi2024bounding}.

The quantum quench in the quantum-enhanced method can propose configurations far from the previous state in Hamming distance. Ideally, this evolution causes the system to dephase within an energy window around the initial state's energy, as then the proposed configurations are likely to be accepted. Difficult sampling problems, such as the low-temperature SK model, often have many nearly degenerate states. The energy windows can be exponentially small, suggesting the need for long time evolution. The proposal probability can be expanded in terms of the eigenstates of $H$,
\begin{equation}
    Q(x|y) = \sum_{n,m}e^{-i(E_n-E_m)t}\braket{x}{n}\braket{n}{y}\braket{y}{m}\braket{m}{x}.
\end{equation}
In the long time limit, and in the absence of many degenerate gaps, the proposal probability relaxes to its equilibrium value, equal to the time-averaged probability \cite{review_Anatoli}. Therefore, in the above form, only the terms where $n=m$ contribute,
\begin{equation}
Q(x|y) =\sum_n |\braket{x}{n}|^2|\braket{n}{y}|^2.
\label{eq:Q-long-t}
\end{equation}
It can be shown, as presented in Appendix~\ref{appendix:time_average}, that the spectral gap for this time-averaged proposal upper bounds the time-average of the gap. As a result, this limit has shorter mixing times than the average performance of the strategy of randomly selecting $t$ for each MCMC step. Numerical studies of the spectral gap dependence on time, presented in Appendix~\ref{appendix:time_average}, show the best mixing occurs once the system has reached equilibrium. The long-time limit, therefore will be the focus of the analysis. In addition, this regime alleviates the need for fine-tuning the evolution time.

\begin{figure}
    \centering
    \includegraphics[width=0.99\columnwidth]{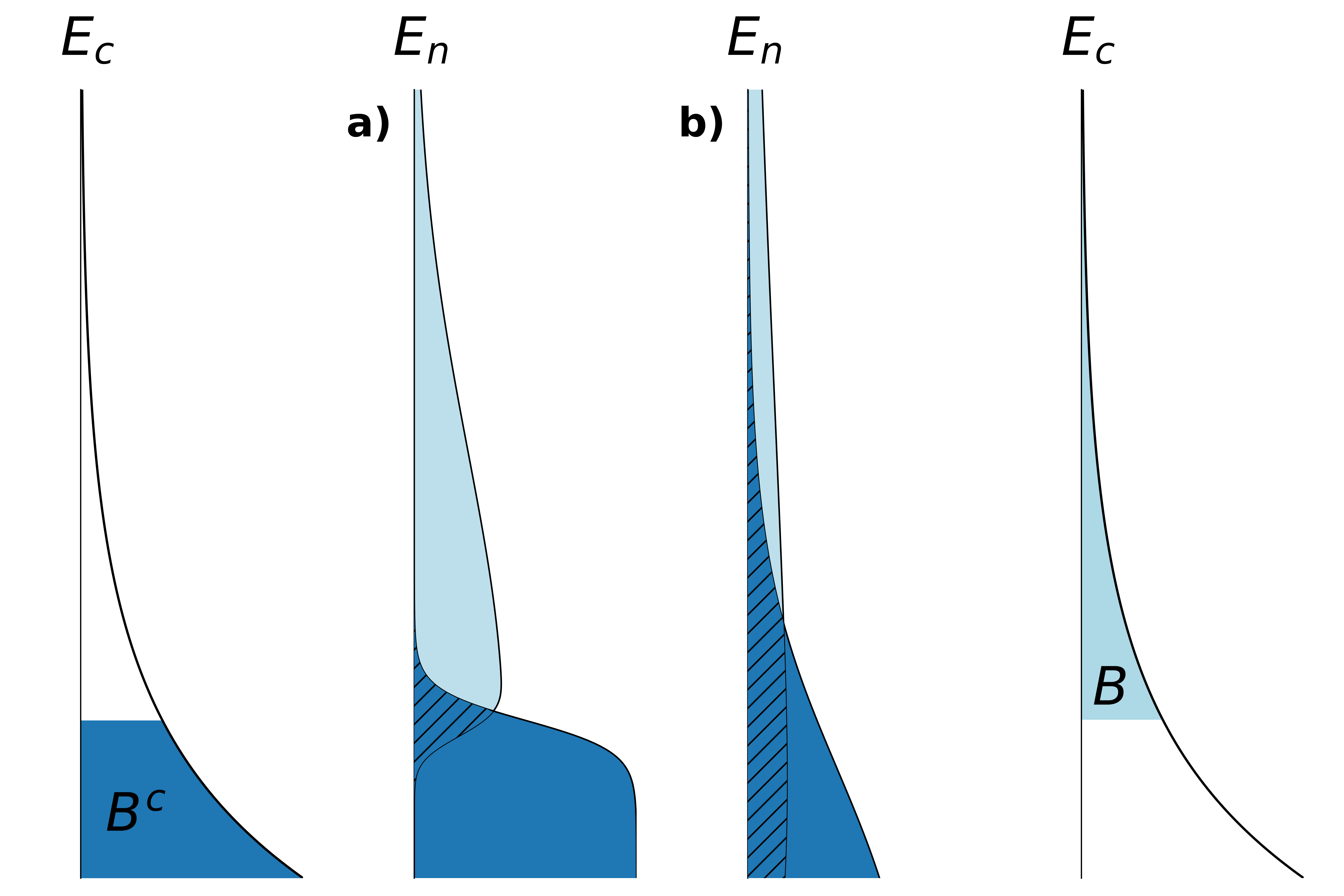}
    \caption{The two right and leftmost plots illustrate the partitioning of the Boltzmann distribution into the set $B$ and its complement. The bottleneck bound of Eq.~\eqref{eq:B-bound} involves the product of two distributions $f(n)$ and $g(n)$, which are shown in light and dark blue, respectively, in the center plots corresponding to two different quantum proposals. In panel a) there is little overlap between the two distributions as the quantum states closely track the classical states, not ideal for fast mixing. In panel b), the overlap is enhanced significantly, but so is the entropy of each distribution, highlighting the balance at play.}
    \label{fig:sets}
\end{figure}

In this limit, the bottleneck upper bound provides valuable insight into the properties of the quantum evolution that lead to favourable mixing. Consider a set $B$ where all configurations in $B$ have higher energy than those in $B^c$. An example of such a set partition within the Boltzmann distribution is illustrated in the left and rightmost panels of Fig.~\ref{fig:sets}. As the Boltzmann measure weighs the equilibrium flow, the bound for this type of set $B$ is typically tight in the low-temperature limit. 
The bottleneck bound takes the following form,
\begin{equation}
    \begin{split}
    \Lambda(B) &= \frac{1}{\pi(B^c)}\sum_n \sum_{x\in B,y\in B_c} \pi_B(x)|\braket{x}{n}|^2|\braket{n}{y}|^2,\\
    &=  \frac{1}{\overline{\pi}(B^c)}\sum_n f(n)g(n),
    \end{split}
    \label{eq:B-bound}
\end{equation}
with $\pi_B(x)$ the Boltzmann measure normalized over $B$. Two distributions are defined,
\begin{equation}
    f(n) = \sum_{x\in B}\pi_B(x)|\braket{x}{n}|^2,
\end{equation}
the probability of transitioning from the set $B$ to the $n$th quantum eigenstate, weighted by $\pi_B$ and 
\begin{equation}
    g(n) = \frac{1}{|B^c|}\sum_{y \in B^c}|\braket{n}{y}|^2,
\end{equation}
the probability that elements of $B^c$ move to the eigenstate $n$. Finally, $\overline{\pi}(B^c)=|B^c|^{-1}\sum_{x \in B^c} \pi(x)$ denotes the average Boltzmann probability to be in set $B^c$. As Eq.~\eqref{eq:B-bound} involves the product of these distributions, one achieves fast mixing if there is significant overlap between them. For example, Fig.~\ref{fig:sets} illustrates these distributions for two quantum proposal strategies that differ by field strength $h$. In the first strategy, shown in Fig.~\ref{fig:sets}(a), the distributions $f(n)$ and $g(n)$ have minimal overlap, resulting in unfavourable mixing compared to in Fig.~\ref{fig:sets}(b), where the overlap is significantly greater. 

The purity of the two eigenstate distributions $f$ and $g$ limits the extent of their overlap and consequently, the spectral gap. This result follows from the  Cauchy-Schwarz inequality applied to Eq.~\eqref{eq:B-bound},
\begin{equation}
    \delta \leq \frac{1}{\overline{\pi}(B^c)} \left(\sum_n f(n)^2 \sum_n g(n)^2\right)^{1/2}.
    \label{eq:gapfgbound}
\end{equation}
These distribution purities can be related to a physical quantity of the quantum evolution, namely the inverse participation ratio (IPR) of classical configurations in the eigenstate basis, defined as,
\begin{equation}
    \mathrm{IPR}(x) = \sum_n |\braket{n}{x}|^4.
    \label{eq:IPR}
\end{equation}
Specifically, applying Cauchy-Schwarz inequality again, the following bound on the gap is obtained,
\begin{equation}
    \delta \leq \frac{1}{\overline{\pi}(B^c)} \sqrt{ \left(\sum_{x\in B} \pi_B(x)\mathrm{IPR}(x) \right) \left(\sum_{y \in B^c} |B^c|^{-1} \mathrm{IPR}(y) \right)}.
    \label{eq:gapIPRbound}
\end{equation}
The IPR is a measure of the localization of the classical configuration $x$ on the eigenstates of $H$. The minimum IPR corresponds to a completely delocalized state, with $\mathrm{IPR}(x)= 1/2^N $, whereas a perfectly localized state gives the maximum IPR value of $1$. If the Hamiltonian under which the system is quenched is fully ergodic, the eigenstates are delocalized, and the spectral gap is upper bounded by $1/2^N$. Specifically, the proposal probability of Eq.~\eqref{eq:Q-long-t} is $ Q(x|y) = 1/2^N$, identical to the strategy of picking configurations uniformly at random. This shows that the quantum quench must be non-ergodic to obtain a scaling advantage over the uniform classical strategy.

The IPR of many-body systems generally scales exponentially with the system size, a consequence of the tensor product structure. For instance, if the states $\ket{n}$ in Eq.~\eqref{eq:IPR} are product states where each qubit is rotated away from the classical z-axis by a small angle $\theta$, the IPR would still scale exponentially as ${\rm IPR} \sim \exp(-2\theta^2 N)$. This immediately highlights the central difficulty in obtaining fast mixing with these quantum-enhanced Markov chains as they suffer from an orthogonality catastrophe, causing all terms within the square root in Eq.~\eqref{eq:gapIPRbound} to be exponentially small. A large IPR can be attained if the eigenstates of the quantum evolution are well localized on the classical configurations, specifically when the transverse field is perturbatively small. In this perturbative limit, there is no quantum advantage, as the quantum evolution can be replicated classically. Moreover, the likelihood of shifting spectral weight from $B$ to $B^c$ decreases with a smaller transverse field. This effect is not reflected in the bound of Eq.~\eqref{eq:gapIPRbound} but with Eq.~\eqref{eq:gapfgbound}.

Additionally, consider the following application of Jensen's inequality,
\begin{equation}
    \overline{\pi}(B^c)=\frac{1}
{|B^c|} \sum_{x \in B^c} \pi(x) \geq e^{-\beta E_c+\beta F_\beta},
\end{equation}
where $E_c=|B_c|^{-1} \sum_{x \in B^c} H_c(x)$ and $F_\beta$ is the free energy of the system at inverse temperature $\beta$. Expression~\eqref{eq:gapfgbound} can thus be further bound by, 
\begin{equation}
    \delta \leq e^{-\beta (F_\beta-F_c)}, 
\end{equation}
where $F_c=E_c-T(S_f+S_g)/2$ and $S_{f,g}$ is the R\'enyi 2-entropy of $f$ and $g$, respectively. This form is reminiscent of classical rate equations, where $F_c$ corresponds to the free energy of the transition state. At low temperatures, the Gibbs free energy $F_\beta$ approaches $E_c$, but the entropy of $f$ and $g$ remains finite, provided the transverse field is finite.

The upper bounds derived here link the mixing time of the quantum-enhanced method to the physical properties of the quantum evolution. Achieving fast mixing requires a balance of competing properties: the distributions $f$ and $g$ must have significant overlap but not too much entropy, as that results in an unfavourably small IPR. Specifically, the low-temperature gap is exponentially small and constrained by the IPR of the low-energy states, allowing regimes of slow mixing to be identified.

\section{Performance}
As with any MCMC method, the performance of the quantum-enhanced method is problem-dependent. For example, on an unstructured marked item problem, a quantum proposal method has no advantage over a uniform classical proposal \cite{orfi2024bounding}. Before investigating more complex spin glass systems, it is insightful to consider a simpler analytically tractable example to elucidate some points made in the previous section.
\begin{figure}
    \centering
    \includegraphics[width=0.99\columnwidth]{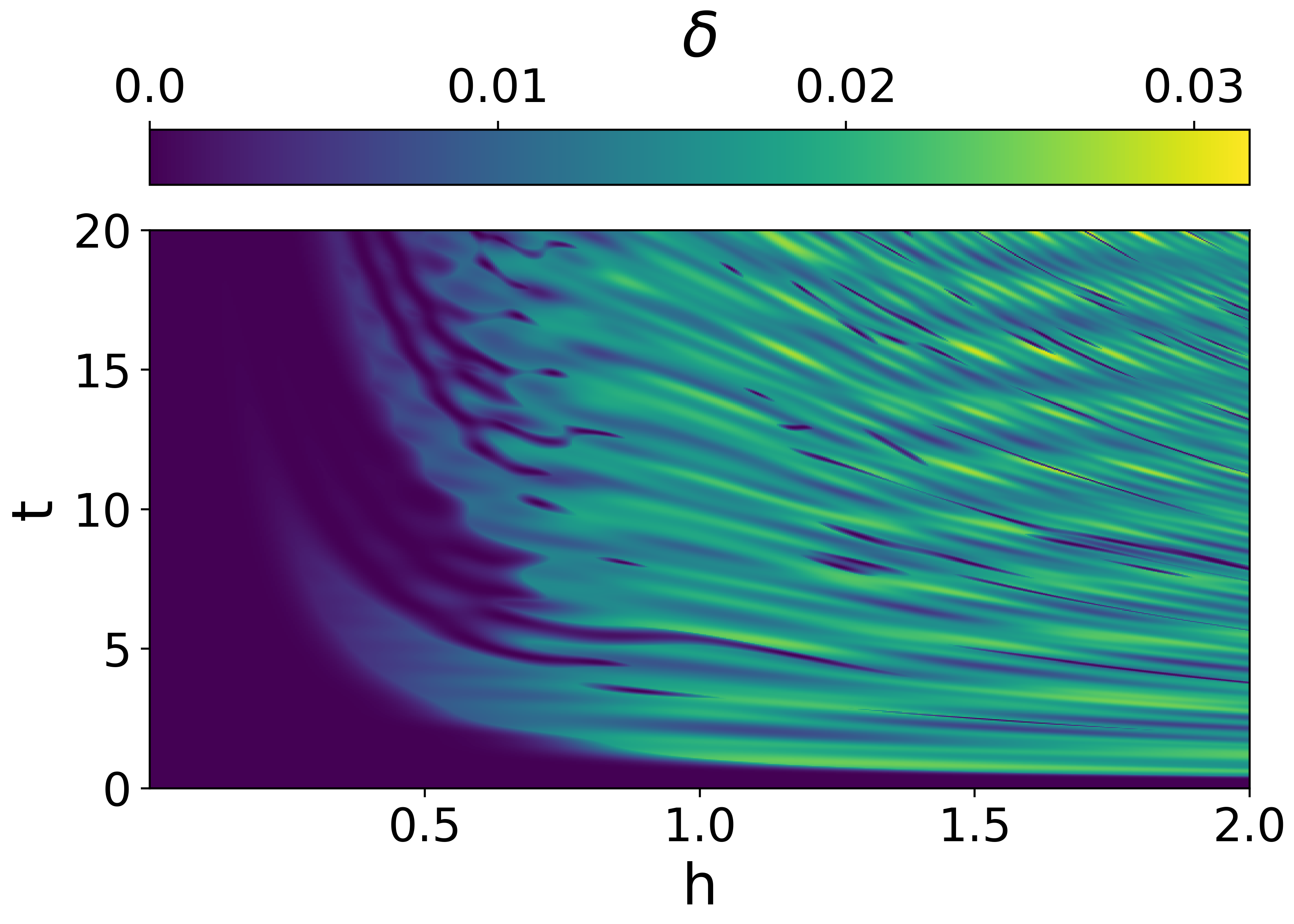}
    \caption{Spectral gap of quantum-enhanced Markov chain Monte Carlo on an eight-site Ising chain at $\beta=5$.}
    \label{fig:ising-grid}
\end{figure}

\subsection{Ising Chain}
Consider the Ising chain with periodic boundary conditions, 
\begin{equation}\label{eq:ising}
    H_c = -\sum_{i=1}^N\sigma_i^z\sigma_{i+1}^z.
\end{equation}
The quantum evolution associated with this system can be solved analytically, allowing the proposal probability to be found for any system size. Unfortunately, this does not mean that the spectral gap can be easily determined, as diagonalizing the transition matrix is feasible only for small systems. As illustrated in Fig.~\ref{fig:ising-grid}, investigations of these small systems show non-trivial dependence on the Hamiltonian parameters $h$ and $t$ but indicate substantial parameter regimes with large spectral gaps. However, it is unclear if this behaviour continues for larger system sizes. Using the bottleneck upper bound, the performance of the quantum-enhanced method on this model can be probed in the thermodynamic limit. 

\begin{figure}
    \centering
    \includegraphics[width=0.95\columnwidth]{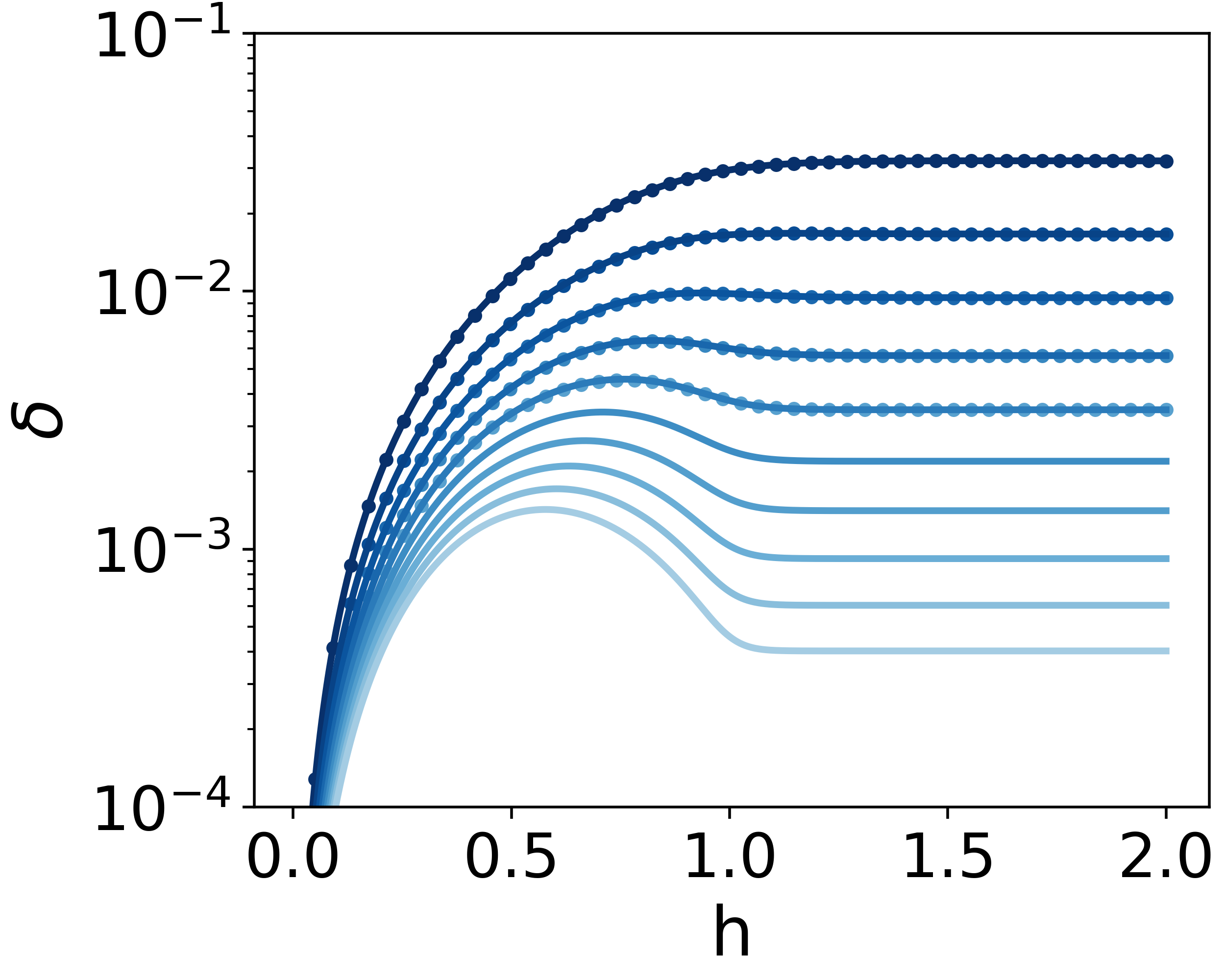}
    \caption{Exact spectral and bottleneck upper bound of the quantum enhanced Markov chain Monte Carlo on the Ising chain at $\beta=5$.  The spectral gap is shown as a scatter plot for even $N$ between 6 and 14, and the solid lines show the upper bound for even $N$ up to 24.}
    \label{fig:ising-bound}
\end{figure}

The possible configuration energies  of the classical system are,
\begin{equation}
    E(k) = -N + 4k,
\end{equation}
where $k=\{0,1,..,N/2\}$. Let $S_k$ be the subset such that $\forall x \in S_k$, $H_c(x) = E(k)$. Consider the choice $B = S_1$ for the upper bound of Eq.~\eqref{eq:lambdaB}. This gives the following bound on the spectral gap,
\begin{equation}
    \delta \leq \frac{1}{N(N-1)}\sum_{x\in S_1,y\in S_0} Q(y|x) + \frac{e^{-4\beta}}{2-N(N-1)e^{-4\beta}},
\end{equation}
where the second term is small in the low-temperature limit. As derived in Appendix~\ref{appendix:ising}, this bound can be expressed in terms of the Hamiltonian parameters $h$ and $t$, 
\begin{equation}
    \delta \leq \frac{2\gamma(h,t)}{N-1}e^{-N\lambda(h,t)} + \frac{e^{-4\beta}}{2-N(N-1)e^{-4\beta}}.
\end{equation}
Where the polynomial prefactor is,
\begin{equation}
     \gamma(h,t) = \int_0^\pi\frac{dk}{2\pi}\frac{h^2\sin^2(k)\sin^2(2t\varepsilon_{k})}{\varepsilon^2_{k}- h^2\sin^2(2t\varepsilon_{k})\sin^2(k)},
\end{equation}
and exponential term, 
\begin{equation}
   \lambda(h,t)= -\int_0^\pi\frac{dk}{2\pi}\ln \left[ 1-\frac{h^2\sin^2(k)}{\varepsilon_k^2}\sin^2(2t\varepsilon_{k})\right],
\end{equation}
with $\varepsilon_k = \sqrt{(h-\cos(k))^2+\sin^2(k)}$. This bound is tight compared to the spectral gap calculated through exact diagonalization on small systems sizes, as seen in Fig.~\ref{fig:ising-bound}, and reproduces the intricate structure seen in Fig.~\ref{fig:ising-grid}. This result confirms our two main points: (i) the gap remains exponentially small unless $\lambda$ vanishes, which only happens in the $h \rightarrow 0$ limit, and (ii) the numerics for small system sizes, which indicate a broad parameter range for speedup, are misleading. In the thermodynamic limit, these features will disappear, and the gap will scale exponentially.

It is a general feature of the quantum-enhanced Markov chain Monte Carlo method that a local classical strategy is reproduced in the small $h$ limit. The quantum proposal can be expanded perturbatively for small $h$, where the proposal probability is only non-negligible for configurations separated by single spin flips. For the Ising chain the local spin-flip strategy is efficient; the spectral gap scales polynomially with the system's size~\cite{wild2021long}. This polynomial scaling is reproduced in the quantum-enhanced MCMC method at small $h$ where the polynomial decay dominates compared as compared to the exponential decay. To leading order $\lambda(h)\sim O(h^2)$ therefore $h \leq O(1/\sqrt{N})$ is needed to avoid exponential scaling. Similarly, $\gamma(h)\sim O(h^2)$ such that the gap increases with $h$ simply because the matrix elements increase. Consequently, the optimal point is around $h_{\rm opt}\sim 1/\sqrt{N}$ and the gap scales like $1/N^2$, in correspondence with the classical result. Hence, similar to the worst-case problem discussed in Ref.~\cite{orfi2024bounding}, there is no speedup over classical MCMC sampling for the 1D Ising chain.  

\begin{figure*}
    \centering
    \includegraphics[width=0.95\textwidth]{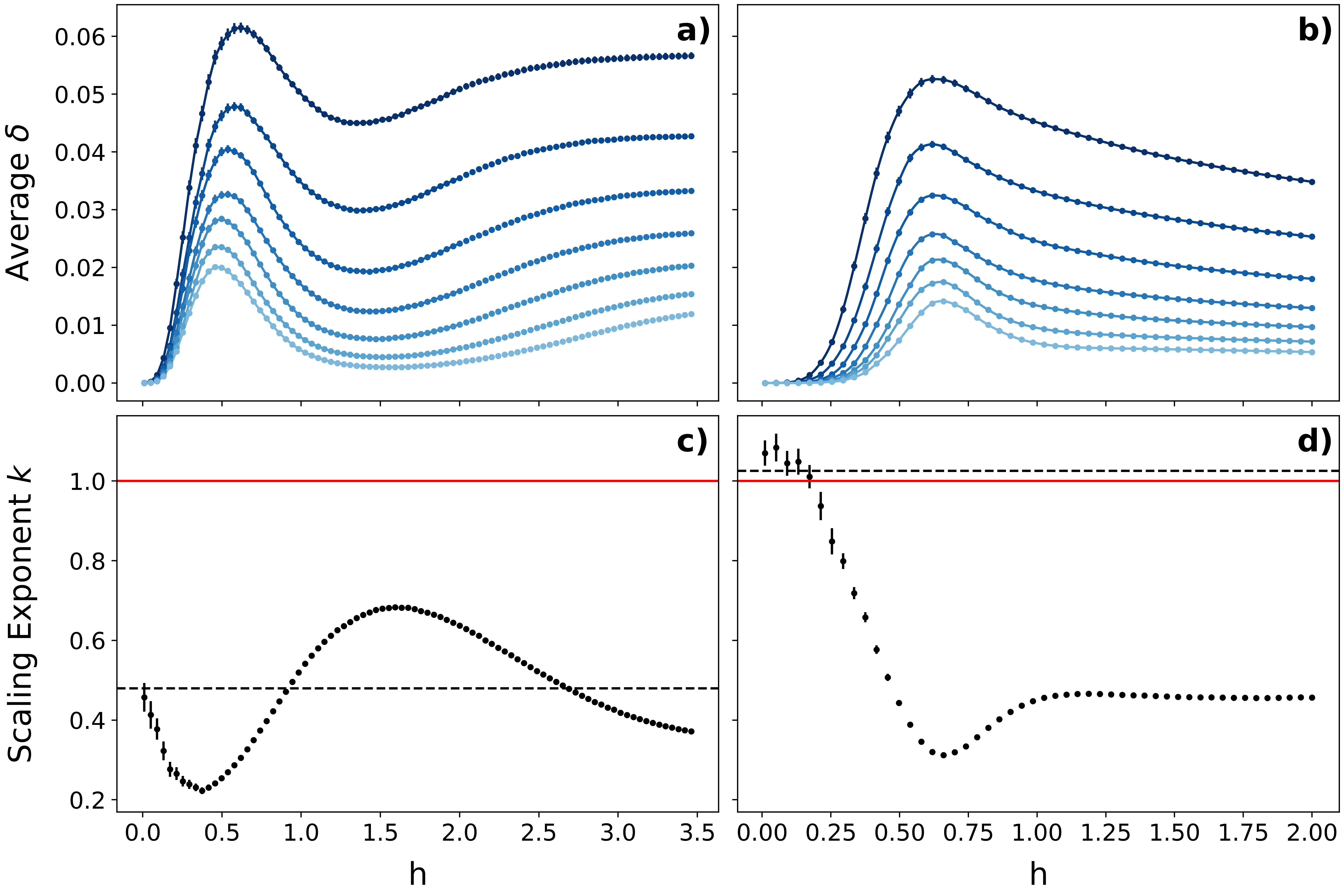}
    \caption{The long-time spectral gap of quantum-enhanced MCMC at $\beta=5$ averaged over 1000 Hamiltonian instances is shown for the Sherrington-Kirkpatrick model (a) and 3-spin model (b) for system sizes $N = 7-13$. The average spectral gap decreases exponentially with the system size,  $\langle \delta\rangle \propto 2^{-kN}$, where the scaling exponent $k$ is displayed in c) and d). The red and dashed black lines indicate the scaling exponent of a uniform classical MCMC method and a modified local classical MCMC strategy.}
    \label{fig:glassGap}
\end{figure*}

\subsection{Spin Glasses}
As the transverse field Ising model is solvable, it was possible to study large system sizes through a bottleneck analysis. However, the Ising model has unique mixing properties that are not relevant to most other systems as it is as easy to sample classically. The main focus of this work is the application of the quantum-enhanced method to spin glass systems, disordered models whose energy landscapes are characterized by many metastable states. Classical chains sampling these spin glass problems at low temperatures are expected to have exponentially slow mixing. Therefore, the optimal transverse field of the quantum-enhanced algorithm does not need to vanish. 

The Sherrington-Kirkpatrick (SK) model, a fully connected spin glass model, is described by the following Hamiltonian,
\begin{equation}\label{eq:SK-ham}
    H_c = -\sum_{1\leq i<j}J_{ij}\sigma_{i}^z\sigma_{j}^z+\sum_ih_i\sigma_i^z,
\end{equation}
with $J_{ij}$ drawn from a Gaussian distribution with a mean of zero and a standard deviation of $1/\sqrt{N}$ \cite{SKmodel}. In the following numerical study, $h_i$ is drawn uniformly between $-0.25$ and $0.25$ to break the $\mathbb{Z}_2$ symmetry. 

Due to Parisi's replica method, the system is known to exhibit a full replica-symmetry breaking (RSB) low-temperature phase, whose energy minima follow a hierarchical clustering structure \cite{spin-glass-phase-mezard1984nature}. The number of these minima is sub-exponential in the system size and separated by large energy barriers, making this phase difficult to sample with a local MCMC method. Finding the ground state of the system, an easier task than sampling, is known to be NP-hard \cite{spinGlassNPhard}. The system has a second-order transition from the replica symmetric paramagnetic phase into this low-temperature spin glass phase.

A generalization of the SK model, known as the p-spin model, has p-wise interactions instead of two-spin interactions \cite{pspin-gardner1985spin},
\begin{equation}
    H_c = -\sum_{1\leq i_1<i_2..<i_p\leq N}J_{i_1i_2...i_p}\sigma_{i_1}^z\sigma_{i_2}^z...\sigma_{i_p}^z.
\end{equation}
The parameters are drawn from a Gaussian distribution with a mean of zero and variance of $p!/(2N^{p-1})$ to ensure the Hamiltonian is extensive. Systems with $ p\geq 3$ do not have only the full-RSB phase of the SK model. Instead, these models exhibit a first-order transition from the paramagnetic phases into a 1-RSB phase. At lower temperatures, there is an additional phase transition into the full-RSB phase \cite{pspin-de1997landscape,pspin-gardner1985spin}. Due to the known differences in these systems, the 3-spin model is also studied.

\begin{figure*}
    \centering
    \includegraphics[width=1 \textwidth]{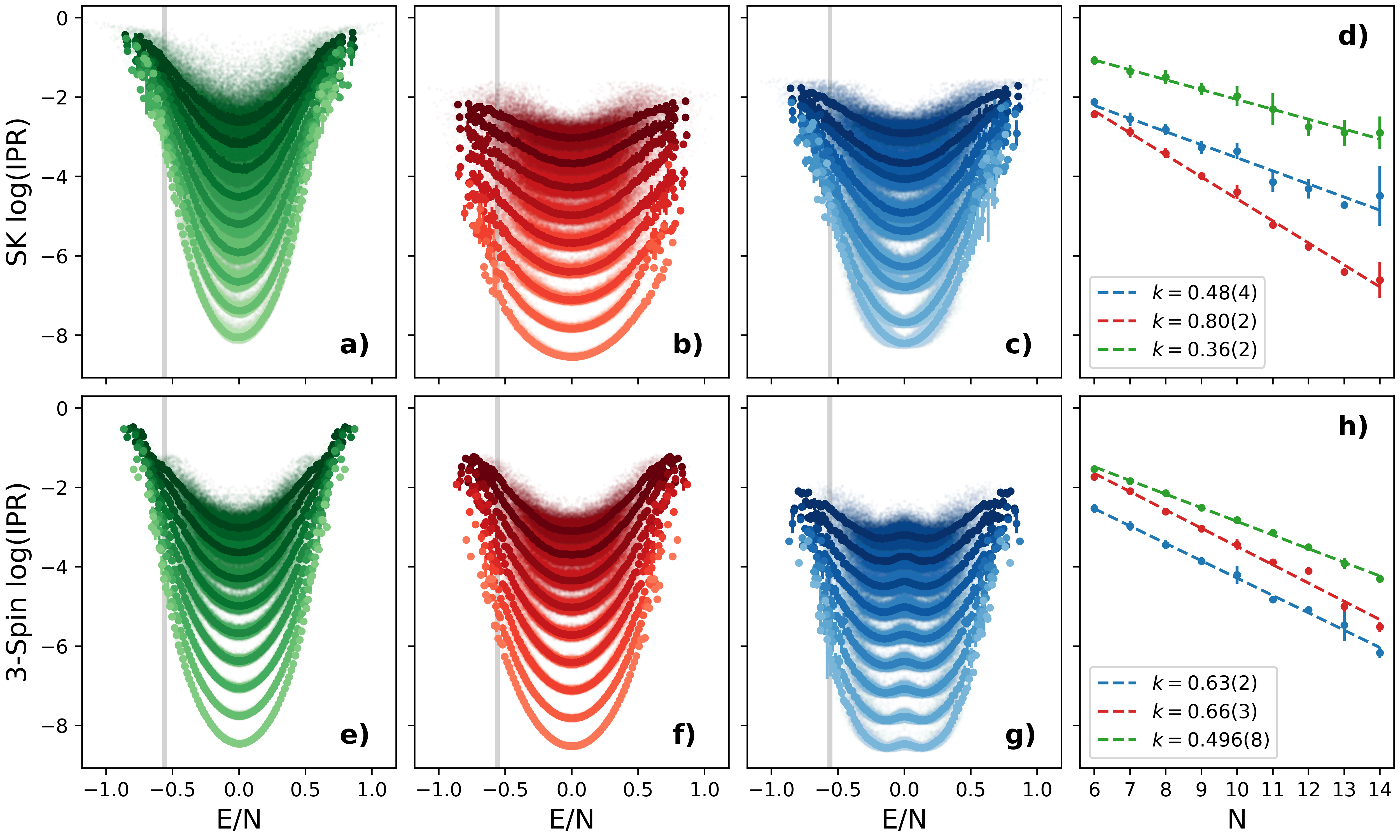}
    \caption{IPR of classical configurations on the quantum evolution Hamiltonian eigenstates for various values of the transverse field $h$. Panels a), b), and c) show the SK model at $h=0.4$, $h=1.6$, and $h=100$, whereas panels e), f), and g) show the 3-spin model at $h=0.7$, $h=1$, and $h=100$. IPR of classical configurations on the quantum evolution Hamiltonian eigenstates for various values of the transverse field $h$. Panels a), b), and c) show the SK model at $h=0.4$, $h=1.6$, and $h=100$, whereas panels e), f), and g) show the 3-spin model at $h=0.7$, $h=1$, and $h=100$. The scaling of the average IPR over an energy window highlighted in grey is shown in panels d) for the SK model and h) for the 3-spin model. The scaling exponent from an exponential fit of the form  $ 2^{-kN}$ shows similar behaviour as the MCMC scaling. }
    \label{fig:IPR}
\end{figure*}

The spectral gap averaged over 1000 Hamiltonian instances is calculated exactly for the SK and 3-spin system in the long time limit, as seen in Fig.~\ref{fig:glassGap} plot a) and b), respectively. The spectral gap scaling with system size determines the algorithm's performance and is the point of comparison to other MCMC methods. In Fig.~\ref{fig:glassGap} plot c) and d), the scaling exponent $k$ is plotted from a least squares exponential fit of the form $2^{-kN}$, on the average gap for each value of $h$. This work focuses on determining the optimal performance regimes of quantum-enhanced MCMC, not an extensive comparison to classical methods. Therefore, the scaling of the spectral gap is exclusively compared to two fixed-temperature classical proposal strategies. 

In the first approach, a new configuration is randomly chosen uniformly, resulting in $ Q(x|y) = 1/2^N$ for all $x$ and $y$. The spectral gap of this method scales inversely with the size of the state space, and the scaling exponent is shown in red in Fig.~\ref{fig:glassGap}. The second method is a local approach where a new configuration is proposed by randomly flipping one spin. In this case, $ Q(x|y) = 1/N$, for $x$ and $y$ separated by a single spin flip, and zero otherwise. The quantum-enhanced MCMC reproduces a local strategy in the small transverse field limit, with proposal probability found from a perturbative expansion to the lowest order in $h$,
\begin{equation}
    Q(x_i|x_j) = \frac{2h^2}{(E_i-E_j)^2},
    \label{eq:localQ}
\end{equation}
where $x_i$ and $x_j$ differ by a single spin flip, say at spin $k$. This probability is independent of $N$, unlike the classical local strategy. For example for the SK model,
\begin{equation}
    E_i-E_j = \sum_{i=1}^N 2J_{ik}+2h_k
\end{equation}
is independent of $N$ due to the central limit theorem. As a result, plotted as the dashed line in Fig.~\ref{fig:glassGap} is the scaling exponent of the spectral gap of the local classical strategy multiplied by $N$. This modified classical scaling agrees with the scaling of the quantum-enhanced MCMC in the small $h$ limit. Unlike the Ising chain, these spin glass problems are known to have long mixing times for local strategies so the spectral gap scales poorly in this limit. The local strategy, gives much better scaling for the SK model than the 3-spin model, indicating the differences in difficulty between these two sampling tasks. For these disordered models, isoenergetic cluster updates are often more successful than local strategies \cite{cluster-zhu2015efficient,houdayer2001cluster}. Additionally, methods such as simulated annealing and parallel tempering can enhance the efficiency of MCMC methods \cite{earl2005parallel}. Therefore, these classical scaling exponents are used as points of comparison, not as a representation of leading classical approaches. 

Increasing $h$ leads to the proposal of higher-order spin flip configurations and an increasing spectral gap. This improvement continues until the effect of the IPR bound becomes relevant. Thus, there is an intermediate optimal regime for both the SK and 3-spin model where the spectral gap is maximized and the system proposes configurations far in Hamming distance but has yet to become fully ergodic. It is at this point that the quantum-enhanced MCMC has the best scaling, outperforming the uniform and local classical strategies. Past this region, further increasing $h$ causes the system to have a larger fractal dimension and, therefore, worse scaling until the large $h$ limit is reached. It is difficult to know the position of this optimal regime a priori as it depends on the quantum evolution's specific eigenstate properties. 

The variations in scaling with $h$ are due to the different localization properties of the quantum eigenstates on classical configurations. The IPR of these configurations, sorted by their energy for different $h$ values, is depicted in Fig.~\ref{fig:IPR}. Panel a) shows the IPR for the SK model at the point of optimal scaling ($h=0.4$), panel b) at the maximally ergodic point ($h=1.6$), and panel c) represents the large $h$ limit ($h=100$). Similarly, panels e), f), and g) respectively exhibit the IPR of the 3-spin model at $h=0.7$, $h=1$, and $h=100$. Individual IPR values are displayed, and their mean over an energy window is plotted in a darker colour for $N=6-14$. In the bottleneck bound of Eq.~\eqref{eq:upperbound}, the proposal probability is weighted according to the Boltzmann measure. Consequently, in the low-temperature limit, the IPR scaling of low-energy configurations gives physical insight into the scaling of the MCMC method. Panels d) and h) display the average IPR within an energy window, indicated in grey in the other panels, as a function of $N$. Fitting exponentially as $ 2^{-kN}$, the IPR scaling exponent matches the behaviour observed in the spectral gap scaling of Fig.~\ref{fig:glassGap}, verifying the relation of the MCMC scaling to the localization properties. 

In both models, the scaling exponent plateaus in the large $h$ limit, never reaching the fully ergodic scaling. In the limit of large $h$, the proposal probability can once again be expanded perturbatively. It is tempting to think that the proposal probability is uniform, since the large $h$ eigenstates are those of $X=\sum_i \sigma^x_i$. However, the latter is highly degenerate, requiring some care in selecting the proper eigenstates. Specifically for the SK model, using the degenerate perturbation theory we find the eigenstates are those of: 
\begin{equation}
    H = \sum_{i<j}J_{ij}\left(\sigma_i^z\sigma_j^z + \sigma_i^y\sigma_j^y\right)+h\sum_i \sigma^x_i.
\end{equation}
 Note that the two terms in the Hamiltonian commute, the eigenstates are independent of the value of $h$ but some small $h$ is required to avoid the emergency of a $\mathbb{Z}_2$ symmetry. This Hamiltonian, at small $h$, is potentially easier to implement as to maintain the needed degeneracy in the original construction $h$ must have no variations across spins, a requirement that is extremely difficult experimentally. Although not achieving the optimal performance of the quantum proposal technique, this large $h$ regime still exhibits a scaling enhancement over the uniform classical method with a simplified experimental procedure. 

\section{Conclusion}
The performance of the quantum-enhanced MCMC method varies significantly across different systems and parameter regimes. In this work, this behaviour is explained through a bottleneck analysis of the chain, which links MCMC performance to the ergodicity of the quantum Hamiltonian. An upper bound on the spectral gap is established, dependent on the IPR of the classical configurations on the quantum eigenstates. An immediate result of this bound is that the quantum Hamiltonian under which the system is quenched needs to be non-ergodic to have improved scaling over the uniform classical method. 

Since this analysis provides an upper bound on the spectral gap, it does not guarantee an improvement in scaling over classical strategies in the non-ergodic regime. Establishing lower bounds through bottleneck analysis is possible but more challenging for this class of problems. However, the bottleneck upper bound offers valuable physical insights into the properties that lead to favourable mixing allowing the identification of parameter regimes where a speedup is expected. We show that there is a generic behaviour across various models, with optimal performance occurring in the long time limit at an intermediate value of $h$. The exact value of this optimal transverse field depends on the specific localization properties of the model and is typically hard to determine. Due to the degeneracy of the quantum Hamiltonian's transverse field, scaling improvement over the uniform strategy persists into the large $h$ limit. Effective Hamiltonians in this regime can be identified as having similar behaviour but with simpler experimental realization, potentially enabling large system-size demonstrations of this algorithm on today's quantum devices.

\section{Acknowledgments}
The authors are grateful for ongoing support through the Flatiron Institute, a division of the Simons Foundation. D.S. is partially supported by AFOSR (Award no. FA9550-21-1-0236) and NSF (Award no. OAC-2118310). We acknowledge support from DARPA-STTR award (Award No. 140D0422C0035). The authors acknowledge useful discussions with Pooya Ronagh and Raymond Laflamme.

\bibliographystyle{naturemag}
\bibliography{references}

\onecolumngrid
\appendix

\section{Time Averaged Upper Bound}
\label{appendix:time_average}
Consider the matrix $Q$ whose elements describe the proposal probability between configurations. In the long time limit, each proposal probability equilibrates to its time-averaged value. Specifically, let $Q^{t_i}$ be the proposal probability matrix at some time $t_i$. The time-averaged proposal probability matrix is then, 
\begin{equation}
    Q = \frac{1}{n}\sum_{i=1}^n Q^{t_i}.
\end{equation}
The time-averaged $P$ is the average of the transition matrices, $P^{t_i}$, associated with proposal probabilities $Q^{t_i}$,
\begin{equation}
    P=\frac{1}{n}\sum_{i=1}^n P^{t_i}.
\end{equation}
This can be shown by relating elements of $P$ to elements of $P^{t_i}$. By construction, $P^{t_i}(y,x) = Q^{t_i}(x|y)M(x|y)$ and $P^{t_i}(x,x)  = 1-\sum_{y\neq x}Q^{t_i}(x|y)M(x|y)$. The off-diagonal elements of $P$ are therefore, 
\begin{equation}
    P(y,x) = \frac{1}{n}\sum_{i=1}^nQ^{t_i}(x|y)M(x|y) = \frac{1}{n}\sum_{i=1}^n P^{t_i}(y,x).
\end{equation}
The diagonal elements are, 
\begin{equation}
    P(x,x) = 1-\sum_{y\neq x}\frac{1}{n}\sum_{i=1}^n Q^{t_i}(x|y)M(x|y) = \frac{1}{n}\sum_{i=1}^n (1 - \sum_{y\neq x}Q^{t_i}(x|y)M(x|y)) = \frac{1}{n}\sum_{i=1}^n P^{t_i}(x,x).
\end{equation}
The gap of this time-averaged transition matrix $P$ is an upper bound on the averaged gap,
\begin{equation}
    \delta[P]\geq \frac{1}{n}\sum_{i=1}^n\delta[P^{t_i}].
\end{equation}
This claim is equivalent to,
\begin{equation}\label{eq:claim}
    \lambda_2[P] \leq \frac{1}{n}\sum_{i=1}^n\lambda_2[P^{t_i}],
\end{equation}
where $\lambda_2$ denotes the second largest eigenvalue. In order to prove this claim, we can make use of the following decomposition of $P$ \cite{levin_MarkovChainsMixingTime},
\begin{equation}
    P = D_\pi^{-\frac{1}{2}}E D_\pi^{\frac{1}{2}}
\end{equation}
where $D_\pi^{\frac{1}{2}}$ is a diagonal matrix with elements $D_\pi^{\frac{1}{2}}(i,i) = \sqrt{\pi(i)}$. An element of the matrix $E$ is then, 
\begin{equation}
    E(y,x) = \sqrt{\frac{\pi(y)}{\pi(x)}}P(y,x).
\end{equation}
A $P$ satisfies the detailed balance condition $E$ is symmetric. Let $\{\varphi_j\}$ be the set of eigenvectors of $E$, then $\varphi_1 = \sqrt{\pi}$ with corresponding eigenvalue $\lambda_1 = 1$. Also let $f_j = D_\pi^{-\frac{1}{2}}\varphi_j$, then $f_j$ is an eigenvector of $P$,
\begin{equation}
    Pf_j = PD_\pi^{-\frac{1}{2}}\varphi_j = D_\pi^{-\frac{1}{2}}D_\pi^{\frac{1}{2}}PD_\pi^{-\frac{1}{2}}\varphi_j = D_\pi^{-\frac{1}{2}}E\varphi_j = \lambda_j D_\pi^{-\frac{1}{2}}\varphi_j = \lambda_jf_j .
\end{equation}
This decomposition allows the eigenvalues of the symmetric matrices $E$ and $E^{t_i}$ to instead be compared. The claim of Eq.~\ref{eq:claim} is equivalent to, 
\begin{equation}
    \lambda_2[E] \leq \frac{1}{n}\sum_{i=1}^n\lambda_2[E^{t_i}].
\end{equation}
The first eigenvector is the same for all $E^{t_i}$ and $E$, $\varphi_1 = \sqrt{\pi}$. Consider then the matrix,
\begin{equation}
    B = E - \varphi_1^T\varphi_1
\end{equation}
then $\lambda_1[B] = \lambda_2[E]$. As each $E^{t_i}$ has the same first eigenvector, we have,
\begin{equation}
    B = \frac{1}{n}\sum_{i=1}^N B^{t_i}.
\end{equation}
The claim can now be written in terms of the largest eigenvalues of the symmetric matrices $B$ and $B^{t_i}$,
\begin{equation}
    \lambda_1[B] \leq \frac{1}{n}\sum_{i=1}^n\lambda_1[B^{t_i}].
\end{equation}
The claim now follows from the convexity of the largest eigenvalue of symmetric metrics, a direct application of the Courant-Fischer Theorem \cite{horn2012matrix}. Specifically, given two symmetric matrices $A$ and $B$,
\begin{equation}
    \lambda_1[A+B]  \leq \lambda_1[A] + \lambda_1[B].
\end{equation}
This result shows that the average gap when choosing $t$ randomly at each step is smaller than the time-averaged gap. 

\begin{figure}
    \centering
    \includegraphics[width=0.95\columnwidth]{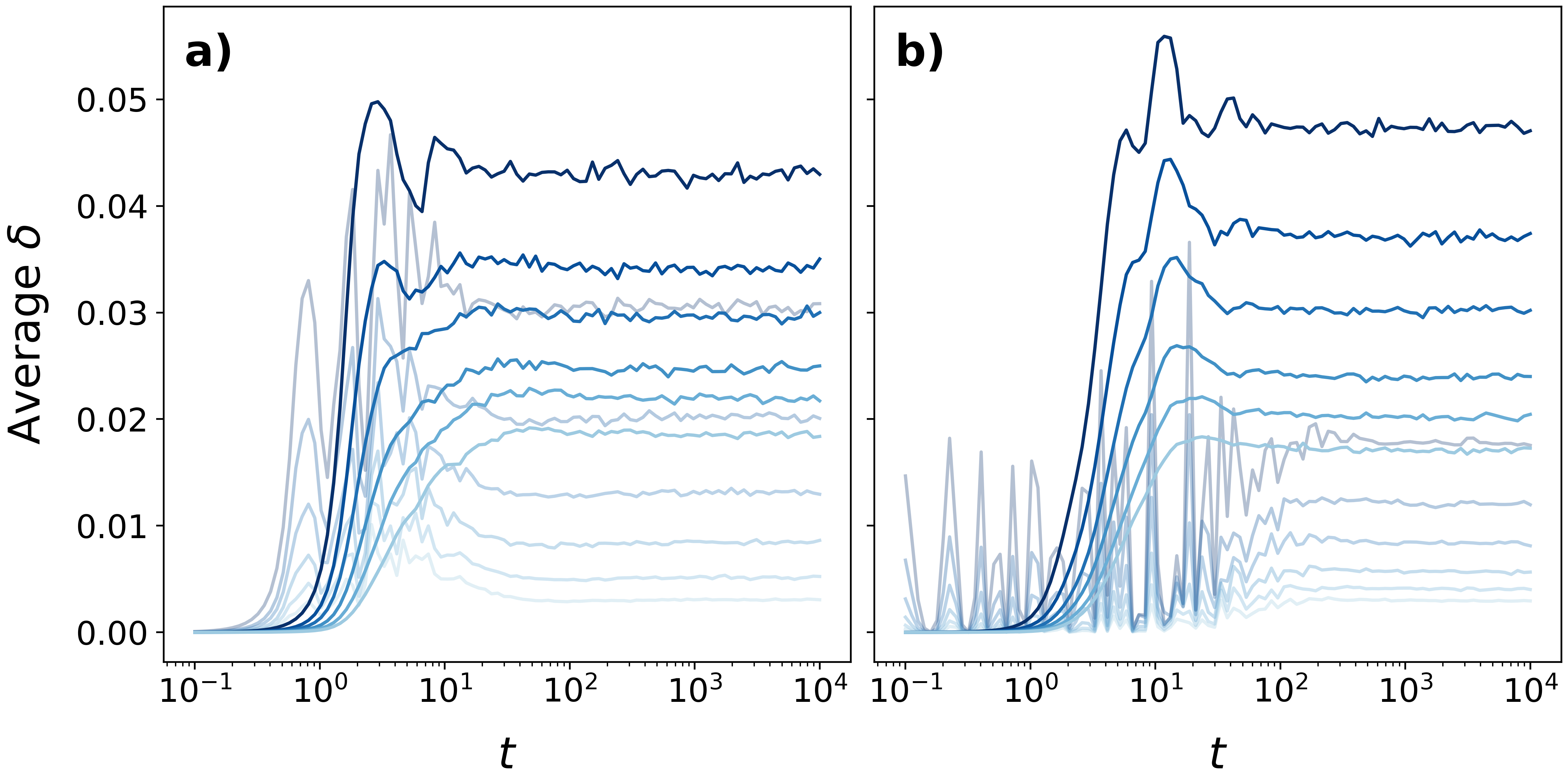}
    \caption{The exact spectral gap averaged over 1000 Hamiltonian instances of the SK model (a) and 3-spin model (b) for system sizes $N=7-12$. The darker and lighter plots represent the time dependence of the average gap at the value of $h$ corresponding to the maximal and minimal gaps, respectively, as seen in the first panel of Fig.~\ref{fig:glassGap}.}
    \label{fig:time}
\end{figure}

Numerical studies of the exact spectral gap of the SK and 3-spin model are shown in a) and b) of Fig.~\ref{fig:time}, respectively. The time dependence is shown for two different values of the transverse field for each model, $h_{\text{max}}$ and $h_{\text{min}}$, corresponding to the maximal and minimal gap values found in Fig.~\ref{fig:glassGap}. Before the system is thermalized, the spectral gap at $h_{\text{max}}$ is below its equilibrium value. For $h_{\text{min}}$, the gap is highly oscillatory, requiring fine-tuning and an understanding of the thermalization properties to take advantage of one of these peak values. These effects further indicate that one must consider the long-time limit to have the optimal behaviour of this algorithm.

\section{Ising Model Upper Bound}
\label{appendix:ising}
For simplicity, we will assume that the number of spins considered $N$, is even. The upper bound of Eq.~\eqref{eq:lambdaB}, where the set $B$ is chosen to be $S_1$ is,
\begin{equation}
    \begin{split}
        \Lambda(B) &= \frac{1}{\pi(S_1)(1-\pi(S_1))}\sum_{x\in S_1,y\in S_1^c}\pi(x)P(x,y),\\
        &= \frac{1}{|S_1|(1-\pi(S_1))}\sum_{x\in S_1}\left[\sum_{y\in S_0}P(x,y) + \sum_{i=2}^{N/2}\sum_{z\in S_i}P(x,z)\right],\\
        &= \frac{1}{|S_1|(1-\pi(S_1))}\sum_{x\in S_1}\left[\sum_{y\in S_0}Q(y|x) + \sum_{j=2}^{N/2}\sum_{z\in S_j}e^{-4j\beta}Q(z|x)\right],\\
        &= \frac{1}{|S_1|}\sum_{x\in S_1,y\in S_0} Q(y|x) + \frac{1}{|S_1|(1-\pi(S_1))}\sum_{x\in S_1}\left[\pi(S_1)\sum_{y\in S_0}Q(y|x) + \sum_{j=2}^{N/2}\sum_{z\in S_j}e^{-4j\beta}Q(z|x)\right].\\
    \end{split}
\end{equation}
The last term can be simplified using the following upper bound, which is tight in the low-temperature regime, 
\begin{equation}
        \pi(B) = \frac{|S_1|e^{-\beta(-N+4)}}{\sum_{k=0}^{N/2}|S_k|e^{-\beta(-N+4k)}} = \frac{|S_1|e^{-4\beta}}{|S_0|+\sum_{k=1}^{N/2}|S_k|e^{-4k\beta}}\leq \frac{|S_1|e^{-4\beta}}{2}.
\end{equation}
Specifically, we then have,
\begin{equation}
    \begin{split}
        &\frac{1}{|S_1|(1-\pi(S_1))}\sum_{x\in S_1}\left[\pi(S_1)\sum_{y\in S_0}Q(y|x) + \sum_{j=2}^{N/2}\sum_{z\in S_j}e^{-4j\beta}Q(z|x)\right]\\ &\leq \frac{e^{-4\beta}}{|S_1|(1-\frac{1}{2}|S_1|e^{-4\beta})}\sum_{x\in S_1}\left[\frac{|S_1|}{2}\sum_{y\in S_0}Q(y|x) + \sum_{j=2}^{N/2}\sum_{z\in S_j}e^{-4j\beta}Q(z|x)\right],\\
        &\leq \frac{e^{-4\beta}}{|S_1|(1-\frac{1}{2}|S_1|e^{-4\beta})}\left[1 + \left(\frac{|S_1}{2}-1\right)\sum_{x\in S_1,y\in S_0}Q(y|x) \right],\\
        &\leq \frac{e^{-4\beta}}{2-|S_1|e^{-4\beta}}.
    \end{split}
\end{equation}
In all, this gives the following upper bound on the spectral gap, 
\begin{equation}\label{eq:ising_bound}
    \delta \leq \frac{1}{N(N-1)}\sum_{x\in S_1,y\in S_0} Q(y|x) + \frac{e^{-4\beta}}{2-N(N-1)e^{-4\beta}}.
\end{equation}
The transverse field Ising model in one dimension is solvable, allowing this bound to be calculated exactly for any system size. Specifically, we require the time evolution of the following Hamiltonian with periodic boundary conditions,
\begin{equation}
    H =  - \sum_{i=1}^N\sigma_i^z\sigma_{i+1}^z+h\sum_{i=1}^N\sigma_i^x.
\end{equation}
This Hamiltonian can be written in terms of fermionic operators through a Jordan-Wigner transformation \cite{sachdevQPT}, 
\begin{equation}
     H = -\sum_i^{N-1}(a_{i}^\dagger-a_i)(a_{i+1}+a_{i+1}^\dagger) - e^{i\pi N} (a_{N}^\dagger-a_N)(a_{1}+a_{1}^\dagger)+ h\sum_i^{N-1} (2a_i^\dagger a_i -1).
\end{equation}
The $\mathbb{Z}_2$ symmetry of the spin problem is now reflected in the fermionic parity $p = \frac{1}{2}(1-e^{i\pi \hat{N}})$. This allows the two parity sectors to be considered independently. The Hamiltonian in these sectors is therefore,
\begin{equation}
     H_p = -\sum_i^{N}(a_{i}^\dagger-a_i)(a_{i+1}+a_{i+1}^\dagger) + h\sum_i^N (2a_i^\dagger a_i -1)
\end{equation}
with the following boundary conditions,
\begin{equation}
a_{N+1} = (-1)^{p+1}a_{1}.
\end{equation}
This Hamiltonian can be further simplified by transforming into momentum space with the following operators,
\begin{equation}
a_k = \frac{1}{\sqrt{N}}\sum_{j=1}^N e^{-ikj}a_j,
\end{equation}
where the different boundary conditions in each parity sector are accounted for by using the following set of k values,
\begin{equation}\label{eq:kvalues}
K_p = \bigg\{k= \pm\frac{2\pi}{N} \times \begin{cases}
(l-1/2) \quad \text{with } l= 1,2,...,N/2 & p=0\\
l \quad \text{with } l= 1,2,...,N/2-1 & p=1
\end{cases}
\end{equation}
This gives a simplified form of the Hamiltonian,
\begin{equation}
    H_p = \sum_{K_p}\left[\left(h-\cos(k)\right)\left(a_k^\dagger a_k-a_{-k}a_{-k}^\dagger\right) +i\sin(k)\left(a_{-k} a_{k}-a_k^\dagger a_{-k}^\dagger\right)\right]\\
\end{equation}
The values of $k$ can be organized into pairs $(k,-k)$ allowing the Hamiltonian to be written as the sum over positive $k$ values, $K_p^+$,
\begin{equation}
    H_{0} = \sum_{K^+_0} H_k, \quad\quad\quad H_1 = \sum_{K^+_1} H_k + H_{k=0,\pi}. 
\end{equation}
Here the $k=0$ and $k=\pi$ terms of the $p=1$ sector have been separated, 
 \begin{equation}
    H_{k=0,\pi} =2(\hat{n}_\pi-\hat{n}_0)+2h(\hat{n}_0+\hat{n}_\pi-1).
\end{equation}
Now each $H_k$ can be in terms of $2\times 2$ matrix $\mathbf{H}_k$,
\begin{equation}
    \begin{split}
    H_k &= 2\left(h-\cos(k)\right)\left(a_k^\dagger a_k-a_{-k}a_{-k}^\dagger\right) +2i\sin(k)\left(a_{-k} a_{k}-a_k^\dagger a_{-k}^\dagger\right),\\
    &=2\begin{pmatrix}
    a_k^\dagger & a_{-k}
    \end{pmatrix}
    \begin{pmatrix}
    h-\cos(k) & -i\sin(k) \\ 
    i\sin(k) & -h+\cos(k)
    \end{pmatrix}
    \begin{pmatrix}
    a_k \\ a_{-k}^\dagger
    \end{pmatrix}, \\
    &=2\begin{pmatrix}
    a_k^\dagger & a_{-k}
    \end{pmatrix}
    \mathbf{H}_k
    \begin{pmatrix}
    a_k \\ a_{-k}^\dagger
    \end{pmatrix}.
    \end{split}
\end{equation}
This matrix can be diagonalized through a Bogoliubov transformation, a mapping to a new set of fermionic
operators through a unitary transform,
\begin{equation}
    \gamma_k = u_k a_k -iv_k a_{-k}^\dagger
\end{equation}
with $u_k = \cos(\theta_k/2)$ and $v_k = \sin(\theta_k/2)$. This defines the transformation matrix, 
\begin{equation}
    A = \begin{pmatrix}
    \cos(\theta_k/2) & -i\sin(\theta_k/2)\\ 
    -i\sin(\theta_k/2) & \cos(\theta_k/2)
    \end{pmatrix},
\end{equation}
where $\theta_k$ is chosen to diagonalize $H_k$. Specifically, take
\begin{equation}
    \begin{split}
        \cos(\theta_k) &= \frac{h - \cos(k)}{\sqrt{(h - \cos(k))^2+\sin^2(k)}},\\
        \sin(\theta_k) &= \frac{\sin(k)}{\sqrt{(h - \cos(k))^2+\sin^2(k)}}.
    \end{split}
\end{equation}
After this transformation, $H_k$ takes on the simplified form, 
\begin{equation}
    \begin{split}
    H_k &= 2\begin{pmatrix}
    a_k^\dagger & a_{-k}
    \end{pmatrix}AA^\dagger
    \mathbf{H}_k AA^\dagger
    \begin{pmatrix}
    a_k \\ a_{-k}^\dagger
    \end{pmatrix}, \\
    &= 2\varepsilon_k(\gamma_k^\dagger\gamma_k-1/2),
    \end{split}
\end{equation}
where the dispersion relation $\varepsilon_k$ is,
\begin{equation}
    \varepsilon_k = \sqrt{(h-\cos(k))^2+\sin^2(k)}.
\end{equation}
The ground state of the system, known as the Bogoliubov vacuum, is the state which annihilates $\gamma_k$ for all $k$. The problem has two bands $\pm \varepsilon_k$, where higher energy states are created by moving fermions from the lower to the upper band. At the phase transition, $h=1$, the two bands collide, allowing zero energy excitations. 

To calculate the bound of Eq.~\ref{eq:ising_bound}, it is easiest to put the classical ground and excited states in terms of these Bogoliubov fermions. Denote the classical ground states as $\ket{y^1},\ket{y^2} \in S_0$, which are the two ferromagnetic states. These states can be decomposed into states with $p=0$ and $p=1$, labelled $\ket{gs_0}$ and $\ket{gs_1}$, respectively, 
\begin{equation}
    \begin{split}
        \ket{y^1} &= \frac{1}{\sqrt{2}}\left(\ket{gs_0}+\ket{gs_1}\right),\\
        \ket{y^2} &= \frac{1}{\sqrt{2}}\left(\ket{gs_0}-\ket{gs_1}\right).\\
    \end{split}
\end{equation}
The classical excited states can be decomposed similarly, with each parity sector containing $N(N-1)/2$ states. As $U$ conserves parity, there is no method to transition between parity sectors during this evolution. This conservation permits these two sectors to be considered independently,
\begin{equation}\label{eq:excitedStateOverlap}
    \sum_{x\in S_1, y\in S_0} |\bra{y}U\ket{x}|^2
     = \sum_{x\in S_1^{p=0}} |\bra{gs_0}U\ket{x}|^2+\sum_{x\in S_1^{p=1}} |\bra{gs_1}U\ket{x}|^2.
\end{equation}
Consider the $p=0$ sector. However, the same procedure can be followed for both sectors.
The ground state $\ket{gs_0} $ is the Bogoliubov vacuum state, $\ket{gs_0} = \prod_{K_0^+} \ket{0}^c_k$, where the transformation $A(\theta_k^c)$ is defined with $h=0$. To calculate the effect of $U$ on this state, $\ket{\emptyset}^c_k$ should be written in terms of the Bogoliubov fermions specified by $h\neq 0$. 
This is done by first undoing the transformation $A(\theta_k^c)$, then changing into the new Bogoliubov fermions, $\gamma_k$, with $A(\theta_k)$,
\begin{equation}
    A(\theta_k)^\dagger A(\theta_k^c) A(\theta_k^c)^\dagger\ket{0}_{k}^c = A(\theta_k)^\dagger A(\theta_k^c)\begin{pmatrix}
        0\\
        1
    \end{pmatrix} =A(\theta_k)^\dagger \begin{pmatrix}
        -i\sin(\theta_k^c/2)\\
        \cos(\theta_k^c/2)
    \end{pmatrix}= \begin{pmatrix}
    -i\sin((\theta_k^c-\theta_k)/2)\\
    \cos((\theta_k^c-\theta_k)/2)
    \end{pmatrix}.
\end{equation}
This state is now in the diagonal basis of our desired evolution. The same method can be used for the classical excited states. Consider the excited state where the fermion at site $k_1$ is moved from the bottom to top band, 
\begin{equation}
    \ket{x}_{k_1} = \ket{1}^c_{k_1}\prod_{k\neq k_1}\ket{0}^c_k.
\end{equation}
The non-vacuum portion of this state $\ket{1}^c_{k_1}$, is transformed as,
\begin{equation}
    A(\theta_k)^\dagger A(\theta_k^c) A(\theta_k^c)^\dagger\ket{1}^c_k = A(\theta_k)^\dagger A(\theta_k^c)
    \begin{pmatrix}
        1\\
        0
    \end{pmatrix}=A(\theta_k)^\dagger 
    \begin{pmatrix}
        \cos(\theta_k^c/2)\\
        -i\sin(\theta_k^c/2)
    \end{pmatrix}=\begin{pmatrix}
    \cos((\theta_k^c-\theta_k)/2)\\
    -i\sin((\theta_k^c-\theta_k)/2)
    \end{pmatrix}.
\end{equation}
First, consider the needed overlap in the $k_1$ block, 
\begin{equation}
    \bra{1}^c_{k_1} U_{k_1}\ket{0}^c_{k_1}
    = i\left(e^{2it\varepsilon_{k_1}}-e^{-2it\varepsilon_{k_1}}\right)\sin((\theta_{k_1}^c-\theta_{k_1})/2)\cos((\theta_{k_1}^c-\theta_{k_1})/2)=-\frac{h\sin(2t\varepsilon_{k_1})\sin(k_1)}{\varepsilon_{k_1}}.
\end{equation}
Therefore the overlap of the entire state is, 
\begin{equation}
    \begin{split}
    |\bra{gs_0}U\ket{x}_{k_1}|^2 &= \prod_{k\neq k_1}\left[ 1-\frac{h^2\sin^2(k)}{\varepsilon_k^2}\sin^2(2t\varepsilon_{k})\right] \left(\frac{h^2\sin^2(2t\varepsilon_{k_1})\sin^2(k_1)}{\varepsilon^2_{k_1}}\right),\\
    &= \prod_{k\in K_0}\left[ 1-\frac{h^2\sin^2(k)}{\varepsilon_k^2}\sin^2(2t\varepsilon_{k})\right]\left(\frac{h^2\sin^2(2t\varepsilon_{k_1})\sin^2(k_1)}{\varepsilon^2_{k_1}- h^2\sin^2(2t\varepsilon_{k_1})\sin^2(k_1)}\right).
    \end{split}
\end{equation}
This method can be used for all first excited states from both sectors, leading to the following,
\begin{equation}
    \begin{split}
         \sum_{x\in S_1, y\in S_0} |\bra{y}U\ket{x}|^2 &= \prod_{k\in K_0}\left[ 1-\frac{h^2\sin^2(k)}{\varepsilon_k^2}\sin^2(2t\varepsilon_{k})\right]\sum_{k \in K_0}\left(\frac{h^2\sin^2(2t\varepsilon_{k})\sin^2(k)}{\varepsilon^2_{k}- h^2\sin^2(2t\varepsilon_{k})\sin^2(k)}\right)\\
        &\quad+ \prod_{k\in K_1}\left[ 1-\frac{h^2\sin^2(k)}{\varepsilon_k^2}\sin^2(2t\varepsilon_{k})\right]\sum_{k \in K_1}\left(\frac{h^2\sin^2(2t\varepsilon_{k})\sin^2(k)}{\varepsilon^2_{k}- h^2\sin^2(2t\varepsilon_{k})\sin^2(k)}\right).
    \end{split}
\end{equation}
In the long time limit considered in the main text, this further simplifies,
\begin{equation}
     \sum_{x\in S_1, y\in S_0} |\bra{y}U\ket{x}|^2 = \prod_{k\in K_0}\left[ 1-\frac{h^2\sin^2(k)}{2\varepsilon_k^2}\right]\sum_{k \in K_0}\left(\frac{h^2\sin^2(k)}{2\varepsilon^2_{k}- h^2\sin^2(k)}\right)+\prod_{k\in K_1}\left[ 1-\frac{h^2\sin^2(k)}{2\varepsilon_k^2}\right]\sum_{k \in K_1}\left(\frac{h^2\sin^2(k)}{2\varepsilon^2_{k}- h^2\sin^2(k)}\right),
\end{equation}
which is the form used in the upper bound seen in Fig.~\ref{fig:ising-bound}. In the continuum limit, $\{K_0\}=\{K_1\}$ and we have for both sectors,
\begin{equation}
   \lambda(h)=-\lim_{N\rightarrow \infty} \frac{1}{N} \sum_k\ln (1-\frac{h^2\sin^2(k)}{2\varepsilon_k^2})= -\int_0^\pi\frac{dk}{2\pi}\ln \left[ 1-\frac{h^2\sin^2(k)}{2\left(1+h^2-2h\cos(k)\right)}\right],
\end{equation}
and
\begin{equation}
     \gamma(h) = \lim_{N\rightarrow \infty} \frac{1}{N} \sum_k \frac{h^2\sin^2(k)}{2\varepsilon^2_{k}- h^2\sin^2(k)} = \int_0^\pi\frac{dk}{2\pi}\frac{h^2\sin^2(k)}{2(1+h^2-2h\cos(k))- h^2\sin^2(k)}.
\end{equation}
This gives the final form of the bottleneck upper bound,
\begin{equation}
    \delta \leq \frac{2\gamma(h)}{N-1}e^{-N\lambda(h)} + \frac{e^{-4\beta}}{2-N(N-1)e^{-4\beta}}.
\end{equation}

\end{document}